\documentclass{ws-ijmpe}

\usepackage{graphicx}
\usepackage{epstopdf}
\begin{document}

\markboth{Anthony R. Timmins for the STAR Collaboration}{Anthony R. Timmins for the STAR Collaboration}

\catchline{}{}{}{}{}

\title{THE CENTRALITY DEPENDENCE OF STRANGE BARYON AND 
MESON PRODUCTION IN CU+CU AND AU+AU FOR 
$\sqrt{s_{NN}}$  = 200 GEV COLLISIONS\
}

\author{Anthony R. Timmins for the STAR Collaboration}

\address{School of Physics and Astronomy, University of Birmingham\\
Birmingham, B15 2TT,
United Kingdom\\
art826@bham.ac.uk}

\maketitle

\begin{history}
\received{(received date)}
\revised{(revised date)}
\end{history}

\begin{abstract}
Transverse momentum spectra of $\Lambda$ and $K^{0}_{S}$ particles are presented for Cu+Cu $\sqrt{s_{NN}}$ = 200 GeV collisions observed at STAR, and compared to Au+Au measurements at the same energy. For both systems, a number of observables are shown to increase at mid-rapidity ($\mid$\emph{y}$\mid$ $<$ 0.5) with increasing centrality. These are the integrated $\Lambda$ and $K^{0}_{S}$ yields, the integrated $\Lambda$ and $K^{0}_{S}$ yields per participating nucleon, and mid-\emph{${p_{T}}$} (1 GeV/c $\rightarrow$ 4.5 GeV/c) $\Lambda$/$K^{0}_{S}$ ratios. The $R_{CP}$ ratio is found to be higher for the $\Lambda$ yields at mid-\emph{${p_{T}}$} compared to the $K^{0}_{S}$ yields for both the Cu+Cu and Au+Au data. In contrast, when similar numbers of participating nucleons are considered for the Cu+Cu and Au+Au data, an indication of increased bulk strangeness production and a higher mid-\emph{${p_{T}}$} (1 GeV/c $\rightarrow$  4.5 GeV/c) $\Lambda/K^{0}_{S}$ ratio are found, for Cu+Cu.
\end{abstract}

\section{Introduction}

Strangeness is readily produced in Relativistic Heavy-Ion collisions and plays an essential role in the characterisation of strongly interacting matter. In particular, anomalies in the relative production of strange baryons and mesons at mid-\emph{${p_{T}}$}, coupled with the suppression at high-\emph{${p_{T}}$} of those particles in central collisions, have provided strong indications that this initial state is partonic.\cite{1} In the first case, coalescence of co-moving partons is the explanation used to account for the relative surplus of baryons to mesons at mid-\emph{${p_{T}}$}. The long predicted attenuation of high momentum partons traversing a quark gluon plasma is used to describe the second case.\cite{2}     

These proceedings report on an investigation into the centrality dependence of these signatures at mid-rapidity ($\mid$\emph{y}$\mid$ $<$ 0.5). Recently produced Cu+Cu data enables a more detailed comparison to the Au+Au data at $\sqrt{s_{NN}}$= 200 GeV.\cite{3} In particular, the geometry of the small system is better defined for Cu+Cu collisions within the Monte Carlo Glauber framework,\cite{4} and Cu+Cu collisions offer a different geometrical make-up compared to Au+Au collisions of a similar system size.
\newpage
\section{Analysis Procedure}

In order to extract $\Lambda$ and $K^{0}_{S}$ yields as a function of transverse momentum, \emph{${p_{T}}$}, and centrality, the STAR Time Projection Chamber (TPC) is utilized to identify both particles via their dominant weak decay. ($\Lambda \rightarrow p^{+}+\pi^{-}, K^{0}_{S} \rightarrow \pi^{+}+\pi^{-}$) from 33 million Cu+Cu events. Their decay products enter the TPC, and a combination of topological, energy loss, kinematic restrictions are placed to ensure the combinatorial background is minimal and linear. These restrictions are varied as a function of \emph{${p_{T}}$} due to the strongly varying background. 
To calculate the reconstruction efficiency, Monte Carlo particles are generated and propagated through a GEANT detector simulation. The generated charge clusters are then embedded into raw data, and the normal reconstruction process is applied. The efficiency is then defined as the ratio of reconstructed particles to the number of generated particles, subject to the previously mentioned restrictions on the raw data.  It was found to depend strongly on \emph{${p_{T}}$}, and weakly on the event-wise multiplicity. Figure 1 shows the corrected spectra for the $\Lambda$ and $K^{0}_{S}$  particles where the calculated efficiencies have been applied to the raw data.
\begin{figure}[h]
\begin{tabular}{c c}
\includegraphics[width = 0.48\textwidth]{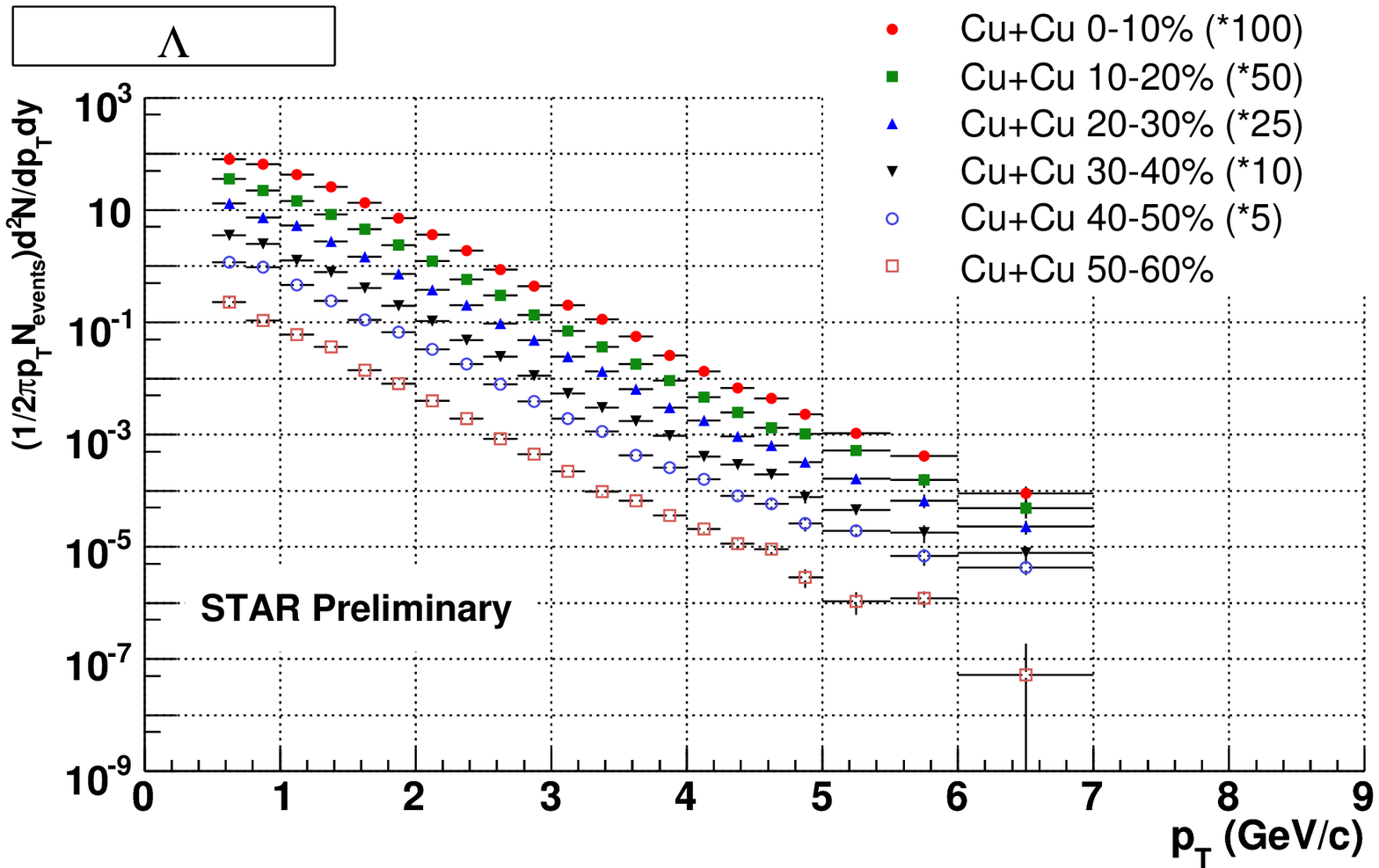}
&
\includegraphics[width = 0.48\textwidth]{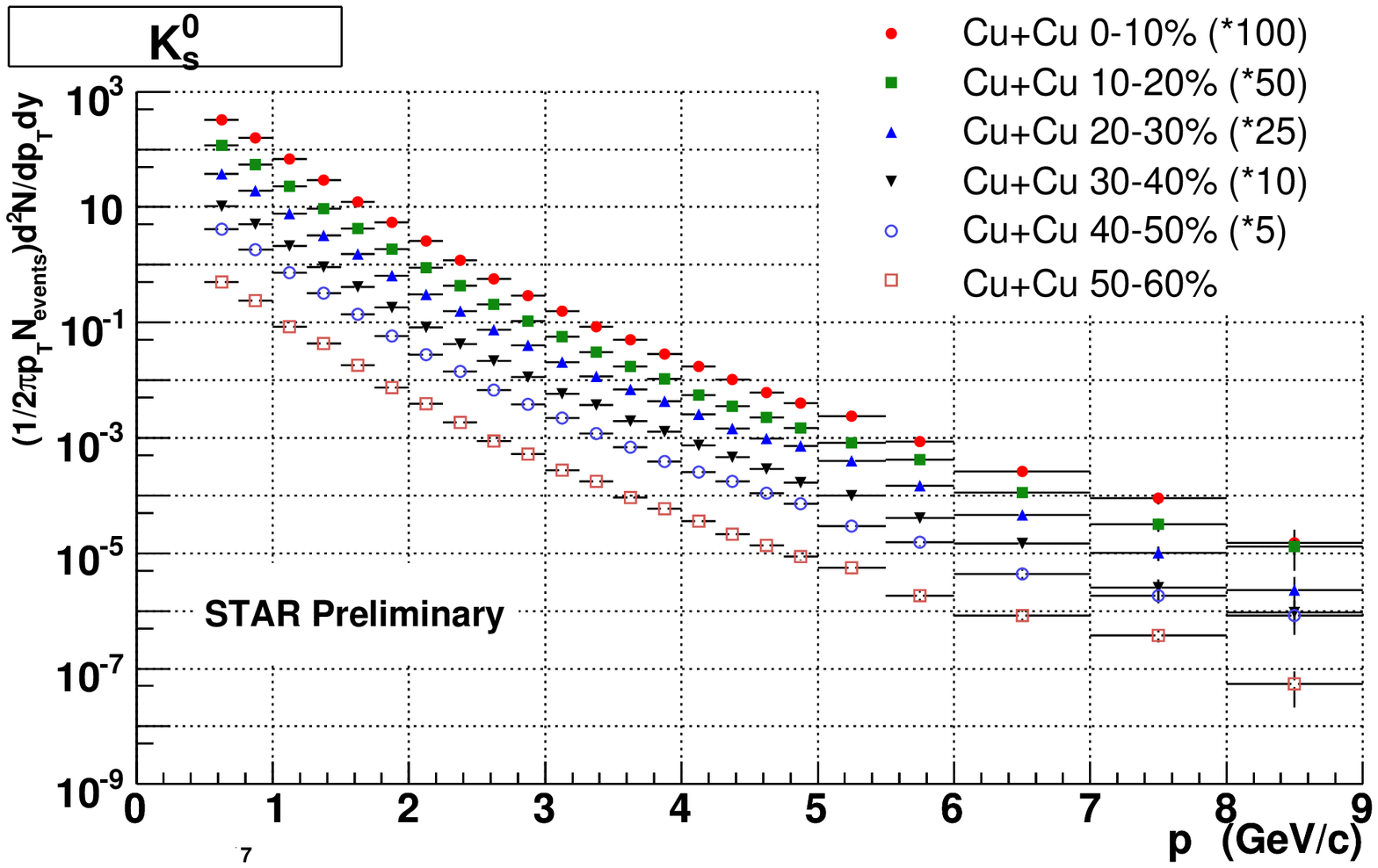}
\end{tabular}
\caption{Corrected $\Lambda$ (left panel) and $K^{0}_{S}$ (right panel) spectra at mid rapidity ($\mid$\emph{y}$\mid$ $<$ 0.5) . The errors are statistical and the $\Lambda$ yields are uncorrected for feed-down.}
\end{figure}
\section{Results}

\subsection{Bulk Yields}

For the integrated Cu+Cu mid rapidity $dN/dy$ measurements shown in figure 2, a Boltzman distribution was used to extract the  $\Lambda$ values, while an exponential distribution was found to be a better fit for the $K^{0}_{S}$ spectra. For both particles, $dN/dy$ appears to increase more rapidly for Cu+Cu as a function of the number of participating nucleons compared to Au+Au.

Figure 2 also shows measurements of $dN/dy$ per $\langle N_{part}\rangle $ which is believed to reflect production per unit volume. The errors are large for these measurements and are dominated by systematic uncertainties in the Glauber calculations. These arise mainly from the uncertainties in the measured cross-section for each collision system and as such, the errors are correlated from point to point for a given system. However, when comparing the two collision systems, the errors will be largely uncorrelated as the uncertainties in the cross sections are independent. The trends indicate increased strangeness production per unit volume for Cu+Cu. This is especially true when the most central Cu+Cu ($\langle N_{part}\rangle\sim 98$) $\Lambda$  and $K^{0}_{S}$ and values are considered. 
\begin{figure}[h]
\begin{center}
\includegraphics[width = 0.65\textwidth]{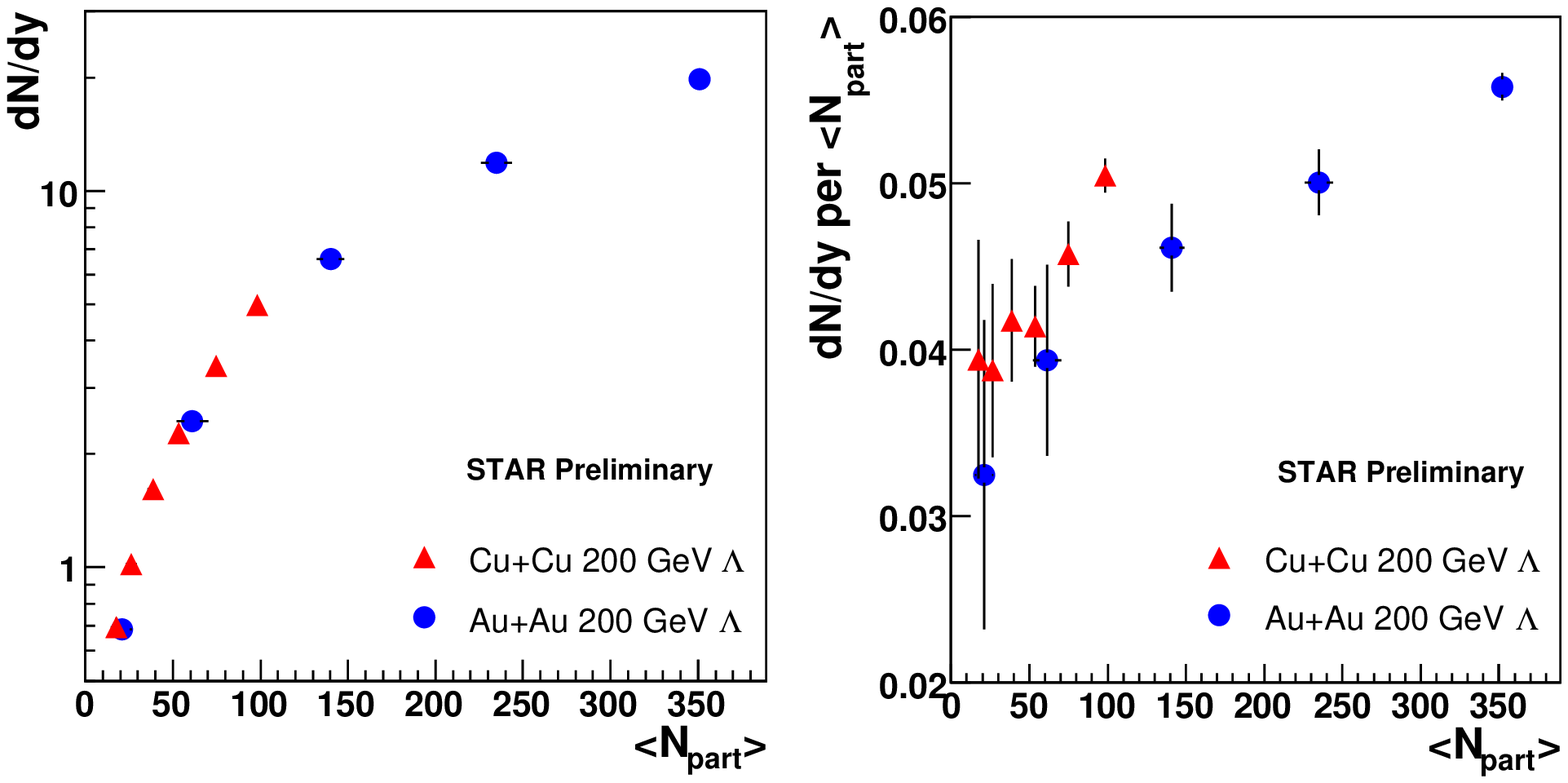}
\includegraphics[width = 0.65\textwidth]{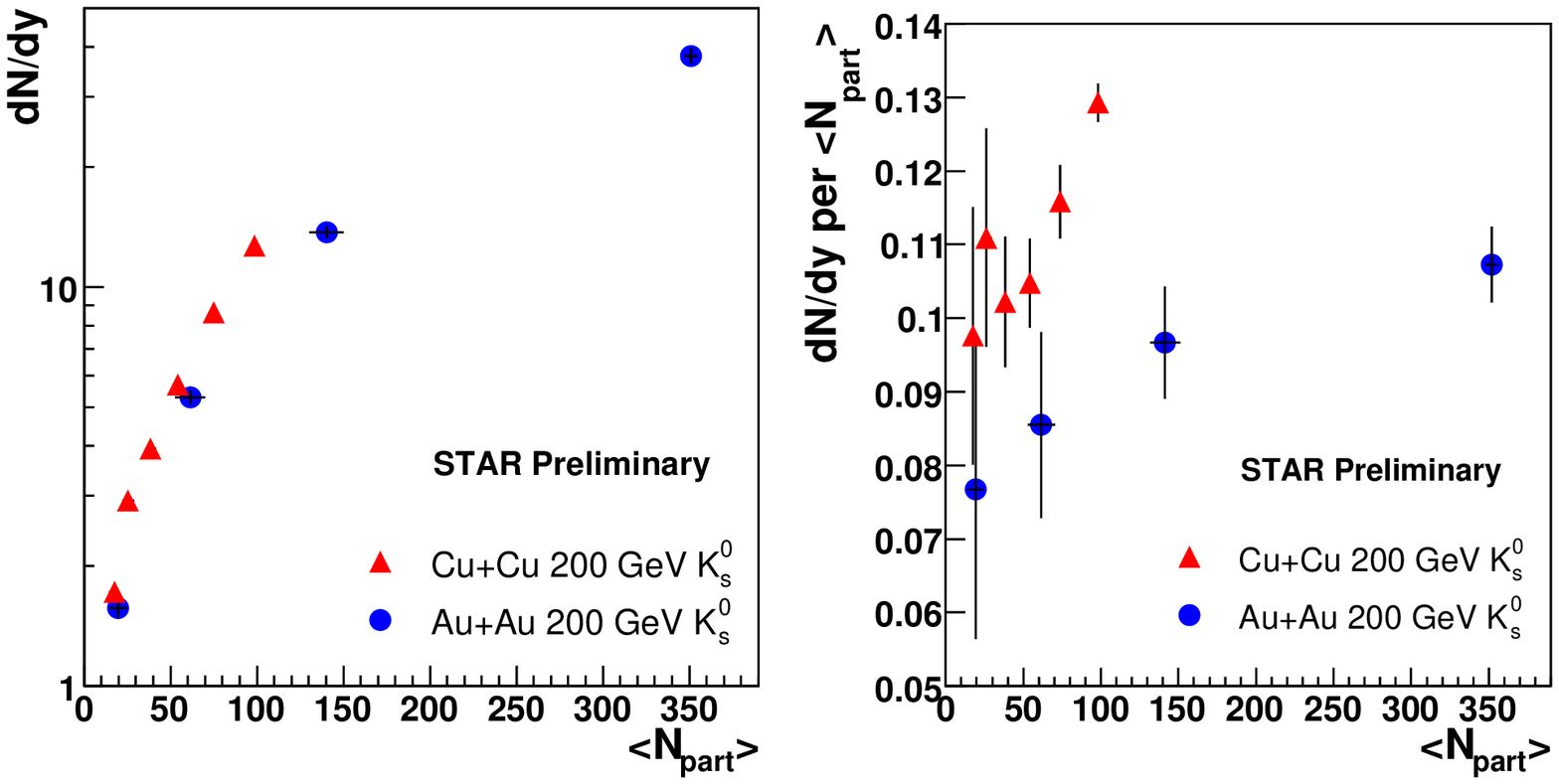}
\end{center}
\caption{$dN/dy$ and $dN/dy$ per $ \langle N_{part} \rangle$ comparisons of $\Lambda$ (top panel) and $K^{0}_{S}$ (bottom panel) for Cu+Cu and Au+Au \protect\cite{5} plotted vs. $\langle N_{part}\rangle$ derived from Glauber model calculations.The errors are statistical and the $\Lambda$ yields are uncorrected for feed-down from weak decays.}
\end{figure}
For lighter systems, strangeness enhancement relative to heavier systems was observed at the AGS and the SPS for similar numbers of participating nucleons. At the AGS, $\langle K^{+}\rangle / \langle N_{part}\rangle $ and $\langle K^{-} \rangle/ \langle N_{part}\rangle $ were significantly higher (factor of 2-3) for $\sqrt{s_{NN}}$  = 5.39 GeV Si+Al and Si+Au collisions compared to $\sqrt{s_{NN}}$ = 5.39 GeV Au+Au collisions\cite{6}. At the SPS, NA49 reported higher $\langle K^{+}\rangle / \langle \pi^{+}\rangle$, $\langle K^{-}\rangle / \langle\pi^{+}\rangle$ and $\langle \phi \rangle/\langle\pi^{+}\rangle$ values for $\sqrt{s_{NN}}$ = 17 GeV Si+Si and C+C collisions compared to Pb+Pb collisions\cite{7} . However, due to different implementations of the Glauber model, care should be taken regarding the definition of $\langle N_{part}\rangle$ as this may vary slightly for different experiments.

\subsection{$\Lambda/K^{0}_{S}$ as a Function of Transverse Momentum}

An inspection of figure 1 indicates two distinct processes with regard to the production of $\Lambda$ and $K^{0}_{S}$ particles with increasing \emph{${p_{T}}$}  for all centralities. Above 2 GeV/c, the trends show an exponential form for the $\Lambda$ yields, and a power-law form for the $K^{0}_{S}$ yields. In an attempt  to compare these processes, figure 3 shows the $\Lambda$/$K^{0}_{S}$ ratio as a function of \emph{${p_{T}}$}  for various centralities in Cu+Cu and Au+Au. In both systems, this ratio is independent of centrality at low-\emph{${p_{T}}$}  ($<$ 1 GeV/c) and high-\emph{${p_{T}}$}  ($>$ 4.5 GeV/c). Each system shows an increase in this ratio with increasing centrality. This trend is observed for $p^{+}/\pi{+}$ and $p^{-}/\pi{-}$ ratios in Au+Au collisions\cite{8}. 
The very fact that this ratio peaks at mid-\emph{${p_{T}}$}  (1 $\rightarrow$ 4.5 GeV/c) may be indicative of partonic recombination mechanism enhancing baryon production relative to meson production for this range. A number of theoretical models have been employed to describe this \cite{9,10,11} and have subsequently been compared to the data\cite{1}. When comparing the magnitude of the peaks for Cu+Cu and Au+Au, again an enhancement is seen for Cu+Cu compared to Au+Au with respect to the number of participants in each collision. It appears that centrality, not $\langle N_{part}\rangle $, offers the more appropriate ordering parameter. 
\begin{figure}[h]
\begin{tabular}{c c}
\includegraphics[width = 0.49\textwidth]{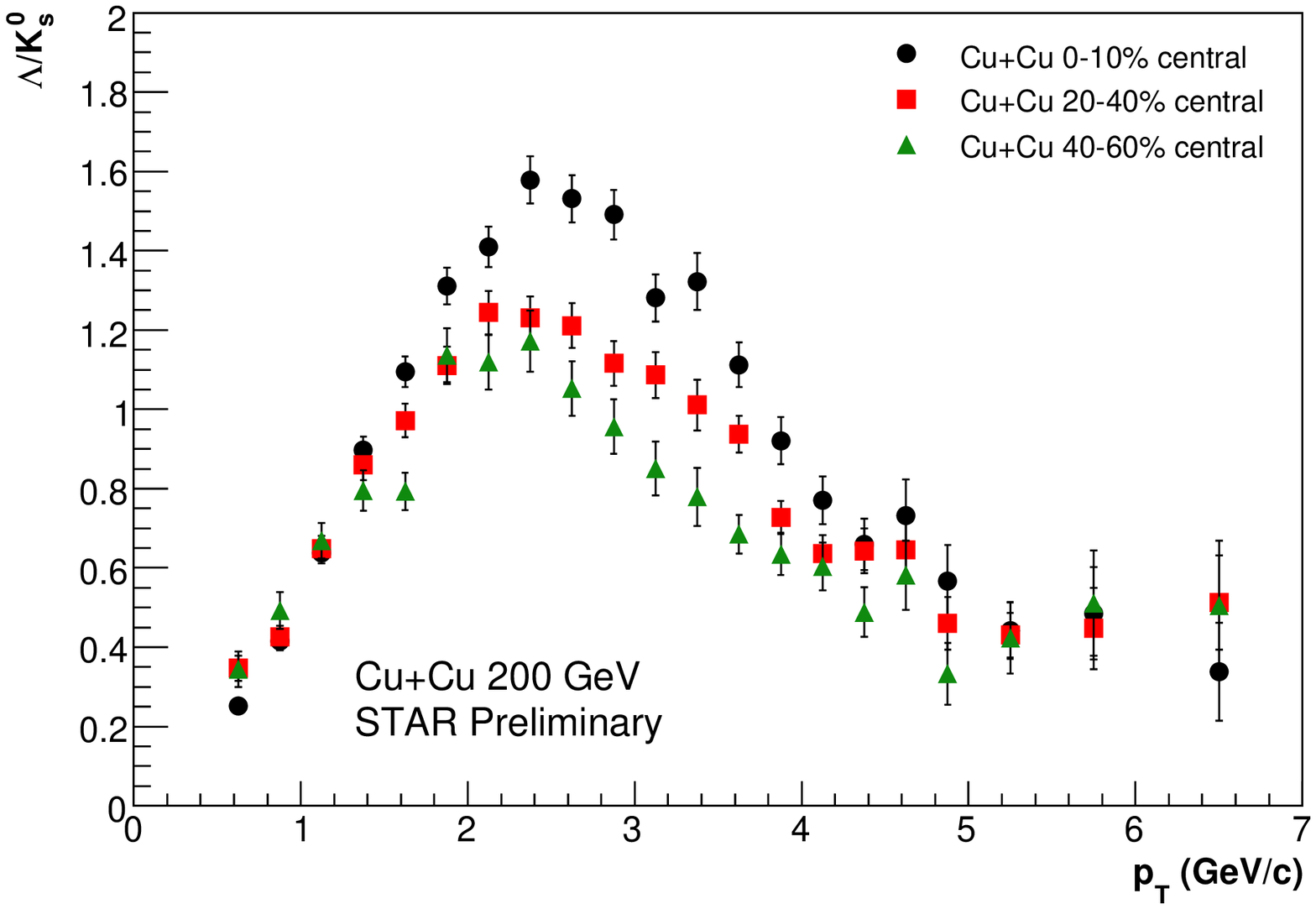}
&
\includegraphics[width = 0.5\textwidth]{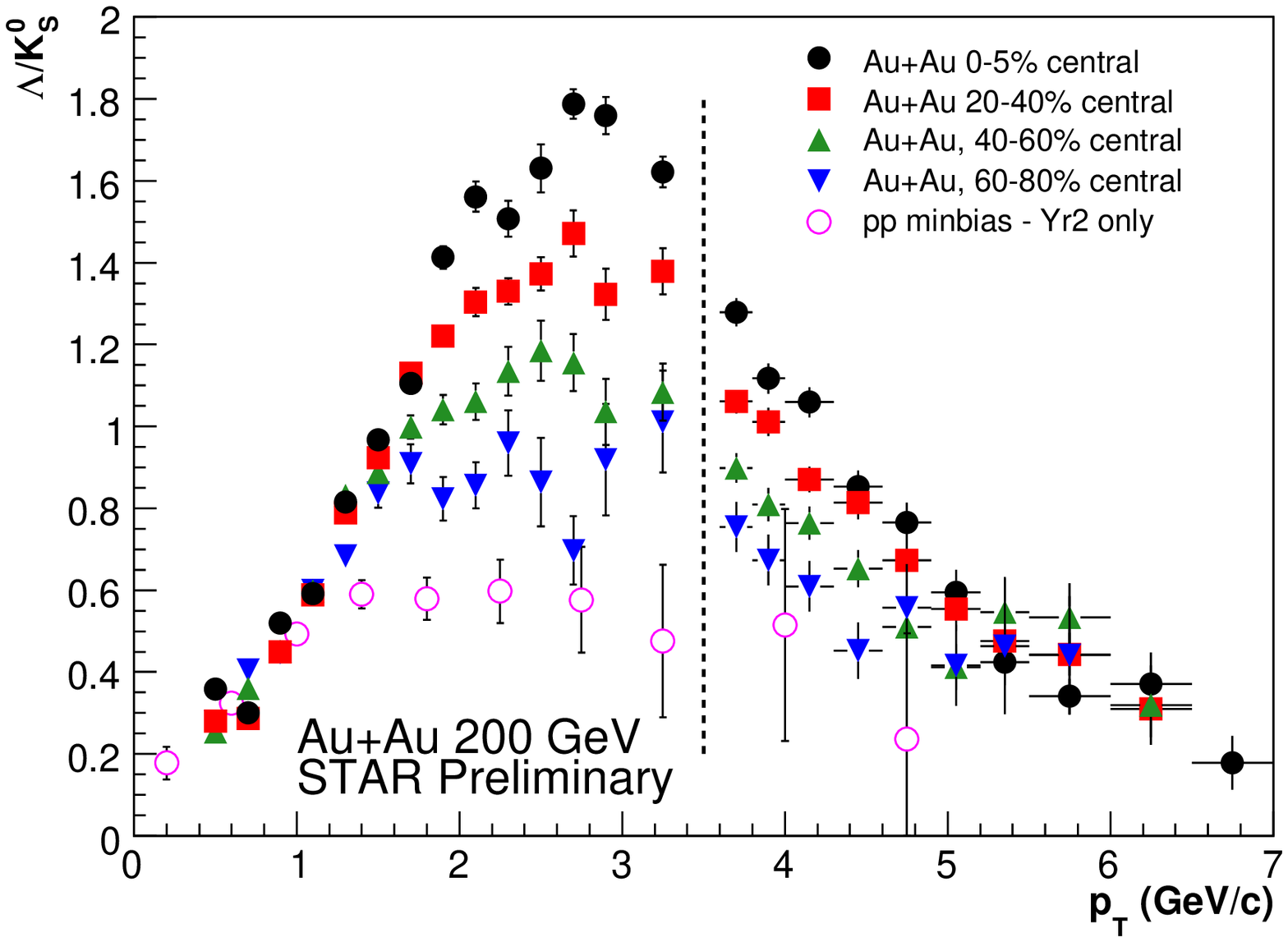}
\end{tabular}
\caption{$\Lambda$/$K^{0}_{S}$ ratio as a function of \emph{${p_{T}}$}  for a) Cu+Cu and b) Au+Au \protect\cite{3}. $\Lambda$ yields are not corrected for feed-down from weak decays. The errors are statistical only. }
\end{figure}
Although large mid-\emph{${p_{T}}$}  baryon/meson ratios in central A+A collisions are often attributed to partonic recombination mechanisms, reasons for these ratios decreasing with centrality are not well understood. This makes understanding any differences in Cu+Cu and Au+Au somewhat more challenging to explain. Clues may be offered when comparing non-strange baryon/meson ratios. BRAHMS reported Cu+Cu mid-\emph{${p_{T}}$}  $p^{+}/\pi^{+}$ ratios consistent with Au+Au for similar numbers of participating nucleons \cite{13}. This suggests that the disparities in the Cu+Cu and Au+Au baryon/meson ratios are confined to strangeness.

\subsection{$R_{cp}$ as a Function of Transverse Momentum}

The processes responsible for the production of strange particles in central and peripheral heavy-ion collisions can also be compared by constructing a ratio ($R_{CP}$) of central yields to peripheral yields, where each yield is normalised by the respective number of binary collisions. Like the number of participants for a given centrality, the number of binary collisions can be derived from Glauber calculations. With increasing \emph{${p_{T}}$}, jet-like processes are expected to increasingly dominate particle production. This should lead to the ratio tending to one in the absence of any medium modification. 
\begin{figure}[h]
\begin{tabular}{c c}
\includegraphics[width = 0.49\textwidth]{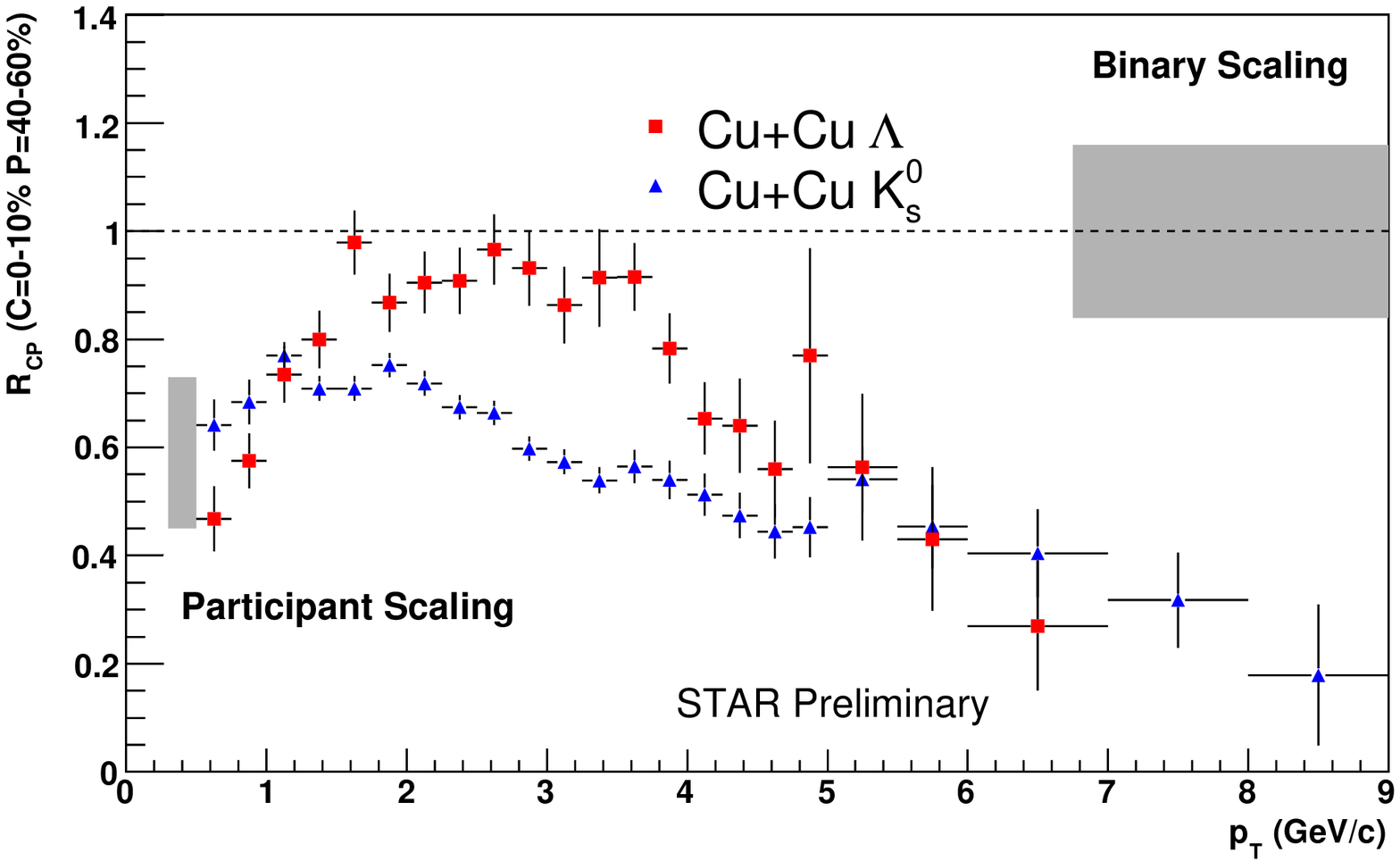}
&
\includegraphics[width = 0.5\textwidth]{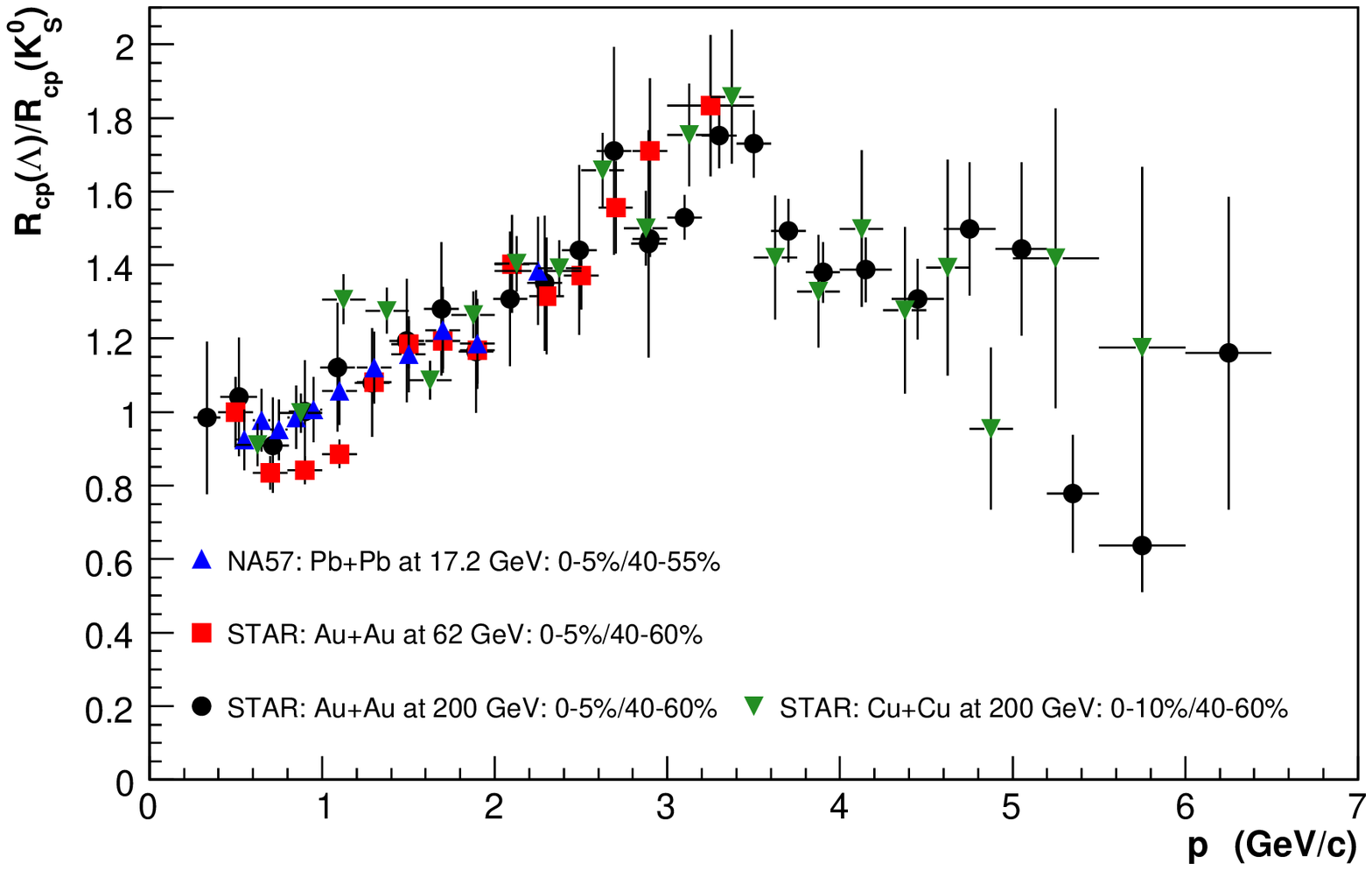}
\end{tabular}
\caption{Left Panel: RCP for $\Lambda$  and $K^{0}_{S}$ in Cu+Cu 200 GeV collisions. The grey bands represent uncertainties in the Glauber calculations for the respective scaling regimes. Right Panel: Double ratio of $\Lambda R_{CP}$ to $K^{0}_{S} R_{CP}$ at a variety of energies and systems \protect\cite{3}. The errors are statistical and the $\Lambda$ yields are uncorrected for feed-down.}
\end{figure}
The left panel of figure 4 shows $R_{CP}$ for $\Lambda$ and $K^{0}_{S}$ yields in Cu+Cu. At low \emph{${p_{T}}$}  ($<$ 1 GeV/c), both particles exhibit participant scaling within the Glauber uncertainties. At mid \emph{${p_{T}}$}  (1 $\rightarrow$  4.5 GeV/c), the respective $R_{CP}$ values diverge as the $\Lambda$ values approach unity and the $K^{0}_{S}$ values steadily decline. For the central data, this can be attributed to recombination playing a larger role in the production of $\Lambda$ particles compared $K^{0}_{S}$ particles in this \emph{${p_{T}}$}  range.  At high \emph{${p_{T}}$}  ($>$ 4.5 GeV/c), the respective $R_{CP}$ values converge and remain below 1 showing a suppression of central $\Lambda$ and $K^{0}_{S}$ yields compared to peripheral $\Lambda$ and $K^{0}_{S}$ yields. This has been observed in Au+Au collisions and is commonly attributed to the suppression of high \emph{${p_{T}}$}  partons in an initially de-confined medium.  
Finally, figure 4 (right panel) shows the double ratio of $\Lambda$ and $K^{0}_{S}$ $R_{CP}$ for approximately the same centrality. For Au+Au and Pb+Pb collisions, there is a remarkable consistency when vastly different collision energies are considered, and this extends to centrality when the smaller Cu+Cu collisions are also examined. This may indicate that the production mechanisms for \emph{${p_{T}}$}  $<$ 3 GeV/c are similar in all the systems represented. 
\section{Conclusions}

We have shown that the production of $\Lambda$ and $K^{0}_{S}$ particles for both Cu+Cu and Au+Au collisions at $\sqrt{s_{NN}}$ = 200 GeV exhibit a number of trends with increasing centrality at mid-rapidity. Regarding bulk production, these are increased yields per participant for both particles. In the case of production at mid-\emph{${p_{T}}$}  (1$\rightarrow$4.5 GeV/c), these are an increasing $\Lambda$/$K^{0}_{S}$ ratio, and a higher RCP value for $\Lambda$ yields compared to the $K^{0}_{S}$ yields. This shows an increasing preference for $\Lambda$ production at mid-\emph{${p_{T}}$}  compared to $K^{0}_{S}$

When comparing both systems with similar numbers of participating nucleons, the integrated $\Lambda$ and $K^{0}_{S}$ yields indicate an enhancement for Cu+Cu collisions relative to Au+Au collisions. In addition, $\Lambda$/$K^{0}_{S}$ ratios are systematically higher at mid-\emph{${p_{T}}$}  for Cu+Cu collisions compared to Au+Au collisions of similar  $\langle N_{part}\rangle $, Whether, in the context of the quark gluon plasma, increased strangeness production at low-\emph{${p_{T}}$}  should be explicitly linked to the relative increased strange baryon production at mid-\emph{${p_{T}}$} , requires further theoretical interpretation. Given that Au+Au mid-\emph{${p_{T}}$}  $\Lambda$/$K^{0}_{S}$ ratios increase with increasing integrated $\Lambda$ and $K^{0}_{S}$ yields per $\langle N_{part}\rangle $, it is perhaps not surprising that when integrated $\Lambda$ and $K^{0}_{S}$ yields per $\langle N_{part}\rangle $ appear enhanced in Cu+Cu (relative to Au+Au), a higher $\Lambda$/$K^{0}_{S}$ ratio is observed. However, it must be stressed that a full determination of systematic uncertainties for each case (Cu+Cu and Au+Au), and the subtraction of the weak decay feed-down contributions to the $\Lambda$ yields, will place all the comparisons in a better context. 

Finally, further measurements from Cu+Cu collisions will shed more light on the nature of strangeness production in heavy-ion collisions. These include: integrated $K^{\pm}, \phi, \Xi^{\pm},\Omega^{\pm}$ yields and respective ratios, and parameters derived from thermal models such as strangeness suppression factor $\gamma_{s}$ and the baryon chemical potential, $\mu_{b}$. As mentioned in section 3.1, at lower energies, strangeness is more readily produced in lighter systems. Indeed, it has been recently predicted that the $\langle K^{+}\rangle / \langle \pi^{+}\rangle$ ratio is $\sim 10\% $ higher for central Cu+Cu collisions compared to Au+Au collisions with similar $\langle N_{part}\rangle $ at RHIC energies \cite{12}. Furthermore, analysis of the Cu+Cu $\sqrt{s_{NN}}$ = 62 GeV data will show further energy dependences regarding strange particle production.

\end{document}